\documentclass[twocolumn,prl,showpacs,preprintnumbers,amsmath,amssymb]{revtex4}
\usepackage{graphicx}
\usepackage{dcolumn}
\usepackage{bm}
\usepackage{color}
\usepackage{ulem}

\begin{document}

\title{Evidence for confined Tamm plasmon modes and application to full control of spontaneous emission }

\author{O. Gazzano}
\altaffiliation{These authors contributed equally to this work}
\affiliation{Laboratoire de Photonique et Nanostructures, LPN/CNRS,
Route de Nozay, 91460 Marcoussis, France}

\author{S. \surname{Michaelis de Vasconcellos}}
\altaffiliation{These authors contributed equally to this work}
\affiliation{Laboratoire de Photonique et Nanostructures, LPN/CNRS,
Route de Nozay, 91460 Marcoussis, France}

\author{K. Gauthron}
\affiliation{Laboratoire de Photonique et Nanostructures, LPN/CNRS,
Route de Nozay, 91460 Marcoussis, France}

\author{C. Symonds}
\affiliation{Laboratoire de Physique de la Mati\`ere Condens\'ee et Nanostructures, Universit\'e Lyon 1, CNRS-UMR 5586, 43 Bd du 11 Novembre 1918, 69622 Villeurbanne Cedex, France}

\author{J. Bloch}
\affiliation{Laboratoire de Photonique et Nanostructures, LPN/CNRS,
Route de Nozay, 91460 Marcoussis, France}

\author{P. Voisin}
\affiliation{Laboratoire de Photonique et Nanostructures, LPN/CNRS,
Route de Nozay, 91460 Marcoussis, France}

\author{J. Bellessa}
\affiliation{Laboratoire de Physique de la Mati\`ere Condens\'ee et Nanostructures, Universit\'e Lyon 1, CNRS-UMR 5586, 43 Bd du 11 Novembre 1918, 69622 Villeurbanne Cedex, France}

\author{A. Lema\^itre}
\affiliation{Laboratoire de Photonique et Nanostructures, LPN/CNRS,
Route de Nozay, 91460 Marcoussis, France}

\author{P. Senellart}
\affiliation{Laboratoire de Photonique et Nanostructures, LPN/CNRS,
Route de Nozay, 91460 Marcoussis, France} 

\date{\today}

\begin{abstract}
We demonstrate  strong confinement of the optical field by depositing  a micron sized metallic disk on a planar interferential mirror.  Zero dimensional Tamm plasmon modes are evidenced both experimentally and theoretically, with a  lateral confinement  limited to the disk area and strong coupling to TE polarized fields. Single quantum dots deterministically coupled to these modes are shown to experience acceleration of their spontaneous emission when spectrally resonant with the mode. For quantum dots spectrally detuned from the confined Tamm Plasmon mode, an inhibition of spontaneous emission by a factor 40$\pm$4 is observed, a record value in the optical domain.
\end{abstract}

\maketitle
Shaping the electromagnetic field around a single emitter leads to acceleration \cite{1} or inhibition\cite{2} of its radiative recombination. 
Full control of the spontaneous emission enables to efficiently collect every single photon generated by a quantum emitter \cite{12} or, reversibly, to ensure that every incident photon interacts with the emitter \cite{13}. These properties are essentials to fabricate bright sources of quantum light \cite{3,4} or a perfect spin-photon interface \cite{5,6}.
Confining the optical field can be obtained by use of total internal reflection or interferences in quasi-periodic structures.  Acceleration of spontaneous emission has been widely demonstrated with various systems like microdisks \cite{14}, micropillars \cite{15} and photonic crystal cavities \cite{16}. 
Purcell effect has also been demonstrated by coupling an emitter to localized surface plasmons, evanescent waves at the interface between a dielectric and a metal \cite{29,30,17}. 

To ensure an optimal coupling between an optical mode and an emitter, inhibition of spontaneous emission can also be very powerful by preventing the emission in any other mode. Yet, inhibition of spontaneous emission is highly demanding on the design of the photonic structure since it requires to cancel the coupling to a continuum of optical modes. Up to now, strong inhibition has only been demonstrated with state of the art photonic crystal structures \cite{7,8} or tapered nanowires \cite{9} with a record inhibition factor of 16.

In this letter, we demonstrate both theoretically and experimentally a new way to confine light in three dimensions of space based on a plasmonic structure, which consists in a micron sized metallic disk  on top of a distributed Bragg reflector (DBR) as illustrated in figure 1a. We show that discrete Tamm plasmon modes appear when decreasing the metallic disk area, with a lateral confinement below the disk area. These mode provide efficient coupling to TE polarized fields and high directionality of emission.  By deterministically coupling single semiconductor quantum dots (QDs) to  confined Tamm Plasmon structures, we demonstrate full control of the spontaneous emission rate over two orders of magnitude, with values as large as 40 $\pm 4$ for the inhibition factor.

In 1931, I. Tamm theoretically showed that localized electronic states can appear at the surface of a crystal due to the breakdown of the periodicity \cite{18}. Recently, Kaliteevski and co-workers have shown an analog confinement of the optical field at the interface between a DBR and 2D metallic layer \cite{10}. The confinement arises, on one side, from the metal's negative dielectric constant, and on the other side, from the DBR band gap. Figure 1b presents the reflectivity of a 40 pairs $\lambda /4n$ GaAs/AlAs DBR for light propagating along z. The mirror stop band is centered at the design energy E$_{DBR}$ =1.42 eV. When a metallic layer is deposited on top of the DBR, a 2D Tamm Plasmon (TP) mode appears as a dip in the reflectivity within the DBR band gap at an energy roughly given by $E_{\text{Tamm}}=\frac{E_{\text{DBR}}}{(1+ \eta E_{\text{DBR}}/E_{\text{plasma}})}$ where $\eta =2 \frac{\vert n_1-n_2\vert}{\pi \sqrt{\epsilon_B}}$ with $n_1$ and $n_2$ being the refractive indices of the DBR layers, $\epsilon_B$ and $E_{\text{plasma}}$ are  the background dielectric constant and the plasma frequency of the metal \cite{10}. With a 50 nm gold layer, the 2D TP mode appears  at $E_{\text{Tamm}}=1.359\ eV$. Figure 1c presents the spatial distribution of the electric field of this mode along z. The optical field maximum is located 40 nm below the semiconductor-metal interface and decays rapidly in the DBR. Without gold, the electric field cannot penetrate inside the DBR layers. This suggests that Tamm plasmon mode could be confined in the three directions of space simply by depositing a gold disk  on a DBR, with lateral diameter of few $\lambda$.

\begin{figure*}
{\includegraphics[width=130mm]{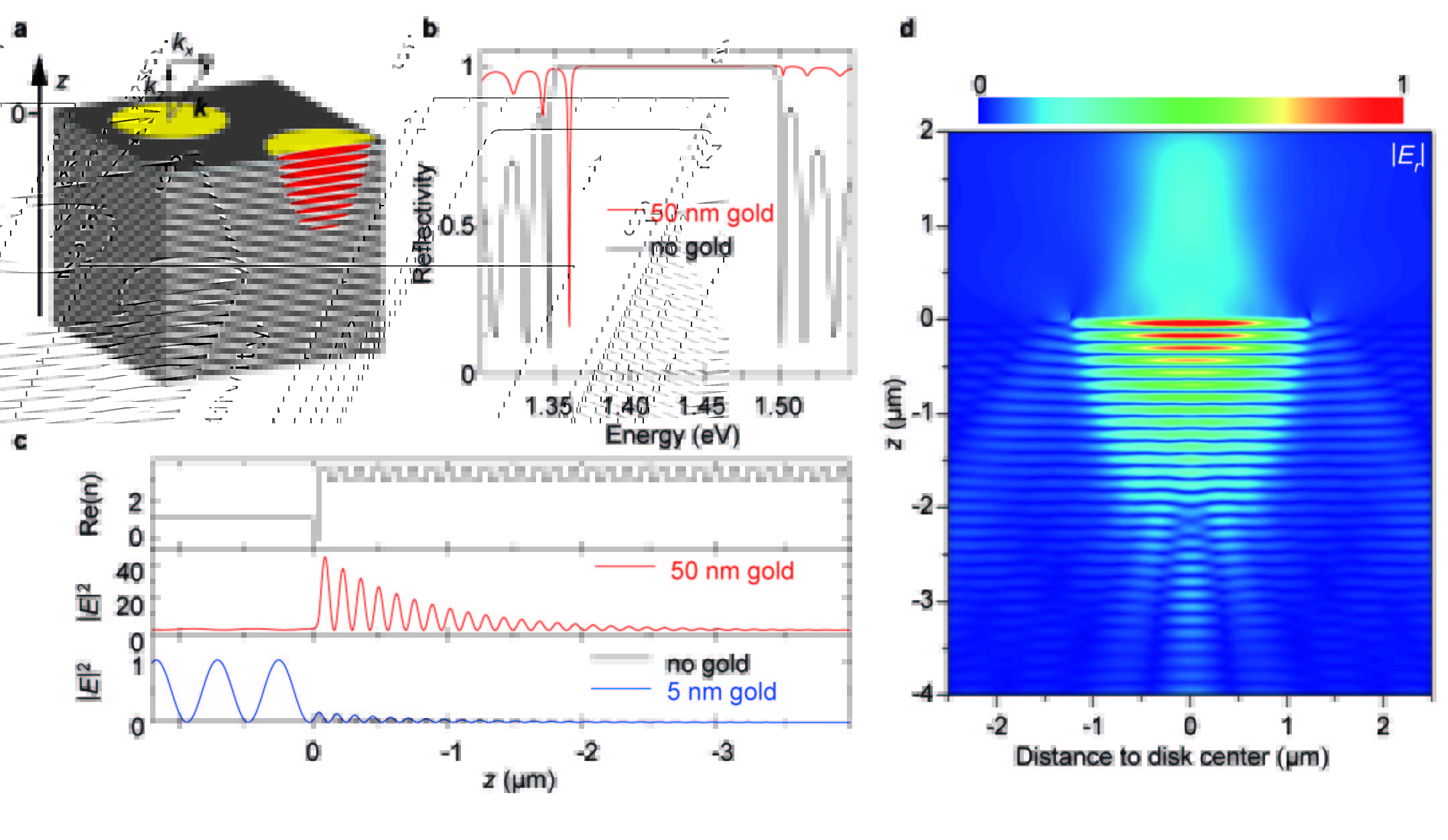}}
\caption{\label{fig:pillarA}a: Schematic of the structure: a 50nm thick gold disk is deposited on top of a planar DBR. The $z$ scale is 1.5 the scale in other $xy$ directions. b, c: 2D Tamm plasmon states b: calculated spectra of a 40 pair GaAs$/$AlAs DBR without (black line) or with (red line) a 50 nm thick gold layer. c: Top: real part of the refractive index along $z$. Below: Spatial distribution of the field intensity along the growth axis for the structure without gold (black), with 50 nm of gold (red) or 5 nm of gold (blue). The mode profile is calculated at E=1.359 eV. d: FDTD simulations: norm of the electric field $\vert E_r \vert$ at the energy of the CTP fundamental mode for a structure with 30 pairs in the DBR and a 2.5 $\mu$m diameter 50 nm thick gold disk. }
\end{figure*}

To experimentally  the existence of these confined TP modes, a DBR with 40 pairs of  $\lambda$/4$n$ layers of Al$_{0.95}$Ga$_{0.05}$As and GaAs is grown by molecular beam epitaxy. A layer of self assembled InAs/GaAs QDs with high QD density  is grown 40 nm below the surface. A 10 nm Al$_{0.1}$Ga$_{0.9}$As layer is inserted between the QDs and the surface. Gold disks  of various diameters are defined using e-beam lithography and gold deposition. No lift-off is performed to prevent collection of emission from QDs outside the disks.    The sample is cooled in a He-flow cryostat and is excited from the top with a laser beam focused onto the sample using a microscope objective. The QD emission is collected by the same objective and is sent to a spectrometer equipped with a charged coupled device. To measure radiation patterns, the Fourier plane is imaged onto the spectrometer slit to measure radiation patterns. Note that the 150 nm thick resist layer does not change the radiation pattern measured with the microscope objective with a 0.55 numerical aperture. 

For this first measurement, a continuous wave  laser at 532 nm is used to excite the QD emission which acts as a white light source filtered by the Tamm structure \cite{19,34}.  Figure 2a shows the measured emission intensity as a function of energy and in-plane wave vector $k_x$ at 40 K. For a diameter of 20 $\mu m$, the parabolic dispersion of the 2D Tamm mode is observed (figure 2a). For diameters below 4 $\mu m$, discrete energy modes are observed, with increasing energy and spectral separation as the size of the disk decreases. Figure 2b shows the angularly integrated spectra. The mode energies, plotted in figure 2c, are well reproduced by a  model similar to the one used to describe micropillar cavities. The modes of strongly confining circular waveguides are vertically confined by the Tamm structure \cite{19}. Radiation pattern measurements also show that the confined Tamm plasmon (CTP) modes exhibit the expected transverse profiles for circular waveguides.    

 The field distribution of the optical modes of the structure are calculated with a finite-difference time-domain (FDTD) method using a freely available software package \cite{27}. The three dimensional system is simulated with a two-dimensional calculation with a cylindrical symmetry. The real part of the gold dielectric constant is fitted to the values reported in ref. \cite{28} using a Drude model, whereas the imaginary part is reduced to account for the experimentally measured Q factor (see below). Perfectly matched layers are used to avoid reflections of light at the boundaries of the analysis space. The electric field associated with the TP mode is calculated after a resonant excitation of the mode. As observed experimentally, the calculations demonstrate that reducing the diameter of the gold microdisk leads to the appearance of discrete energy modes. For each of these modes, strong lateral confinement of the electromagnetic field under the disk area is evidenced. As an example, figure 1d shows the map of the electric field amplitude $\vert E_r \vert$   of the fundamental  CTP mode for a disk diameter of 2.5 $\mu m$.

  The quality factor $Q$ of the modes as a function of disk diameter is plotted in figure 2d. From $Q$ values as high as 1200 observed for large diameters, we deduce a gold refractive index of $n_{\text{gold}}=0.03+5.72i$ indicating the high quality of the gold layer. It is well known that the imaginary part of the dielectric constant depends strongly on the gold deposition method \cite{20} and is  reduced at cryogenic temperatures\cite{21}. For small disks, $Q$ decreases: the spatial distribution of the electric field shows coupling to leaky modes due to scattering at the disk edge (see figure 1d).

As shown in figure 1c, the TP state does not exist for 5-10 nm thin gold layers. It appears only for gold thicknesses larger than 35 nm. The quality factor increases when the gold layer thickness goes from 35 nm to 50 nm and then saturates due to dissipative losses in the gold layer. The 50 nm thickness used here was chosen to optimize both the quality factor and the transmission of the QD emission through the gold layer. We also checked theoretically and experimentally that strong lateral confinement is maintained under a 50 nm thick gold disk surrounded by a 5 nm thick 2D gold layer. In future work, this thin metal layer could be used for electrical contacts to allow for tuning of the QD energy with Stark effect or for electrical injection of charge carriers \cite{26}.

\begin{figure}[t]
\includegraphics[width=85mm]{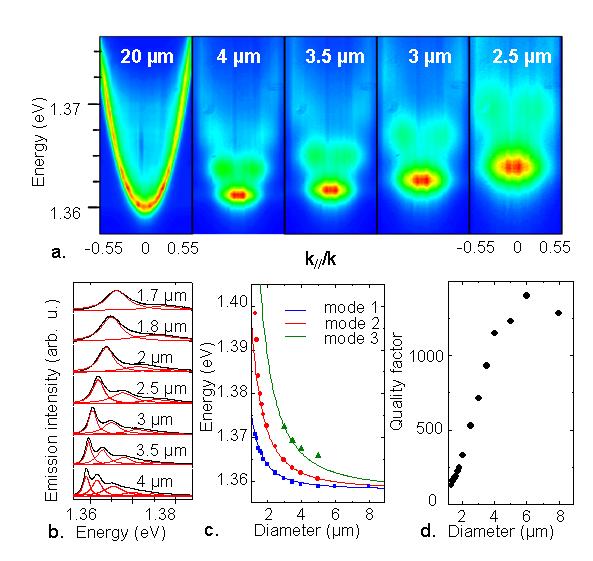}
\caption{\label{fig:pillarB1}a: Measured emission intensity as a function of energy and in-plane wave vector for  various disk diameters. The 2D Tamm dispersion is measured for a disk size of 20 $\mu m$. b: Measured  angularly integrated Photoluminescence PL spectra (black lines). Red lines are Lorentzian fits to the modes. c: Experimental (symbols) and theoretical (solid lines) energies of the CTP modes as a function of the disk diameter. d: Quality factor of the fundamental CTP mode.}
\end{figure}

An emitter ideally coupled to a confined optical mode experiences an acceleration of its spontaneous emission by the Purcell factor, proportional to $Q/V$ where $V$ is the mode effective volume \cite{1}. To investigate the modification of spontaneous emission of a semiconductor QD induced by a CTP mode, we used a sample containing a low density of QDs 40 nm below the surface. First, a 5 nm thin gold layer was deposited over the whole sample surface. Then, we used the in-situ lithography technique \cite{11} to define a 45 nm thick gold disk with its center located between two QDs, spatially separated by approximately 300 nm (figure 3a). The QD2 excitonic (X) transition energy is 6 meV below the QD1 X line. During the insitu lithography step, the power dependence of  both X lines was similar under continuous wave excitation, showing that both lines presented similar radiative lifetimes \cite{33}.
Other lines emerging from QD1 are observed and identified by power dependence and photon correlation spectroscopy: the biexciton XX$_{QD1}$ and the charged exciton CX$_{QD1}$ lines. The disk diameter of 2.6 $\mu$m was chosen so that the CTP mode is spectrally close to the QD1 exciton line at 10 K whereas X$_{QD2}$ is strongly detuned.

 To monitor the modification of spontaneous emission induced by the TP mode, the sample is excited with a pulsed Ti:Sapphire laser (pulse width 1.5 ps) at a wavelength of 850 nm.  Decay times are obtained with a temporal resolution of 300 ps by correlating the QD signal with the laser synchronisation signal  using  Perkin-Elmer photon counting modules. We measure the radiative lifetime of the emission lines X$_{QD1}$ and X$_{QD2}$ as well as the radiative lifetime of several QDs before metal deposition or coupled to a 2D TP state (figure 3c).  QDs before metal deposition have lifetimes of $\tau_{X_{ref}}= 1.3 \pm 0.2$ ns, the same lifetime as in bulk GaAs. However, QDs coupled to the 2D TP mode present decay times around $\tau_{X_{2DTP}} =2.5$ ns. This observation shows that the metal deposition does not introduce non-radiative processes due to defects at the interface which would shorten the carrier lifetime. In contrast, this longer decay time is the evidence for a slight inhibition of spontaneous emission which has been predicted for emitters in sub-wavelength metal or dielectric planar cavities \cite{23,31,32}.

\begin{figure*}[t]
\includegraphics[width=170mm]{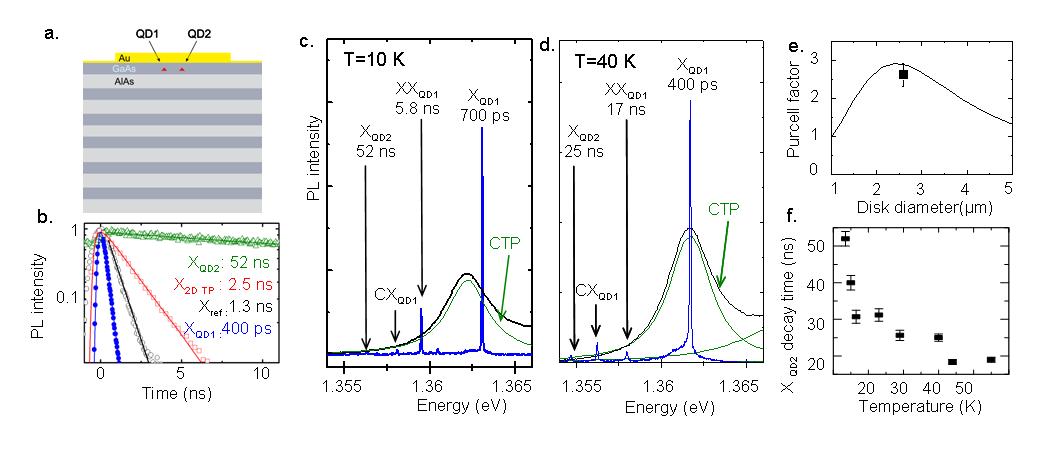}
\caption{\label{fig:Summary}a: Structure of the device: a 50 nm thick gold disk is deposited on top of two QDs (QD1 and QD2). The disk is surrounded by a 5 nm thick gold layer. The disk diameter is 2.5 $\mu$m. b: Blue line: PL spectrum recorded from the device. Black line: PL spectrum recorded on a disk with diameter 2.6 $\mu$m containing a high QD density. The green lines are Lorentzian fit to the CTP modes. c: Decay of the PL recorded at 40K for the X$_{QD1}$  line (full circles), at 13  K, for a QD exciton in bulk sample (open circles), for a QD exciton coupled to a 2D TP mode (X$_{2D TP}$) (open squares) and for the X$_{QD2}$ line (open triangles). Solid lines are  mono-exponential decay fit curves. d: line: Expected Purcell factor deduced from the Q factor measured in figure 2.d. Symbol, measured Purcell factor for the QD1 X line. e: Decay time of the emission of the QD2 X line as a function of temperature.}
\end{figure*}

 At 10 K, X$_{QD1}$ is slightly detuned from the CTP mode (figure 3c) and presents a lifetime of $700 \pm 100 ps$. At 40 K, the lifetime of the X$_{QD1}$ line in resonance to the CTP mode is measured to be as short as $\tau_{X_{QD1}}=400$ ps corresponding to an acceleration of spontaneous emission by a Purcell factor of $F_P=2.5$. Figure 3d presents the theoretically expected Purcell factor deduced from the quality factor measured in figure 2d. Transfer matrix modelling of the electric field along z show that the effective volume of the CTP mode amounts to 40 $\%$ of the mode volume of a pillar microcavity with equal diameter. The acceleration of spontaneous emission observed for X$_{QD1}$ is close to the expected value $F_P=2.9$, for a 2.6 $\mu$m 0D Tamm state with Q=490, showing the good spatial matching between QD1 and the CTP mode. The maximum Purcell factor provided by CTP modes is limited by the quality factor, which is in turn limited by the absorption in the metal. It amounts to 3 for the gold based TP mode under consideration and could be improved with less dissipative metals like silver. 

We now consider the biexciton line from the same QD1, which is detuned on the low energy side both at 10 K and 40 K (figures 3c,d). For this line XX$_{QD1}$, very long decays are observed showing a strong inhibition of spontaneous emission. This inhibition is larger at 40 K (lifetime $17 \pm 3$ ns)  when the XX$_{QD1}$ line is more detuned from the CTP than at 10 K (lifetime 5.8 $\pm 0.3 $ ns). This inhibition can qualitatively be understood considering a recent theoretical work calculating the emission property of a quantum emitter in the vicinity of a metallic disk \cite{24}. When spectrally detuned to from the CTP mode, the  emitter couples to surface plasmons confined at the semiconductor/gold disk interface which should lead to a redirection of its emission close to the disk normal incidence. However, in a CTP structure, emission perpendicular to the disk is forbidden by the DBR stop band. Further theoretical modelling is needed to confirm this interpretation. 

Strong inhibition of spontaneous emission is also observed on other CTP devices when the QD emission line is detuned from the fundamental CTP mode on the low energy side. Typical decay times between 7 ns and 30 ns are observed, depending on the device and the temperature. The largest decays time is observed on the X$_{QD2}$ at 10 K (figure 3b-c): a decay time as long as $\tau_{X_{QD2}}=52$ ns is observed. Compared to the radiative lifetime of similar X lines for QDs before metal deposition, $\tau_{X_{QD2}}=52$ ns corresponds to an inhibition factor  $\gamma^{-1}= \tau_{X_{QD2}}/ \tau_{X_{ref}}$ of 40$\pm$4. To the best of our knowledge, this value is the largest reported value in the optical domain.

Experimental evidence for  strong inhibition of radiative emission are often limited by non-radiative processes which shorten the lifetime of the carriers trapped in the QD \cite{7,8,9}. 
The measured decay time $\tau_{X_{QD2}}$=52 ns show that these non radiative processes are strongly inefficient at 10 K in our structure. This decay time is a lower limit  for the non radiative decay time  at 13K  from which we  deduce a quantum efficiency for the QD larger than $\tau_{X_{QD2}}/(\tau_{X_{QD2}}+ \tau_{X_{ref}})=0.97$. This value is quite remarkable for a QD located 40 nm away from a semiconductor/metal interface. When increasing temperature, a larger inhibition factor is theoretically expected for the X$_{QD2}$ line. However, we observe a shorter decay time, due to thermally activated non-radiative processes (Figure 3f). This observation shows that the inhibition factor of 40 measured at 10 K is indeed a minimum value.

Controlling the interaction between a single quantum emitter and the electromagnetic field, is an important step to use quantum dots for quantum information processing. In particular, the strong inhibition of spontaneous emission provided by CTP modes could increase significantly the time scale for coherent optical and electrical manipulations of the exciton qubit \cite{25}. Combination of acceleration and inhibition enables to fabricate very bright source of quantum light or to implement quantum gate with a single spin in a cavity \cite{5}. Remarkably, the fraction of  the QD emission emitted into the CTP mode $\beta=F_p/(F_p+\gamma) $ is   very close to $1$ since $\gamma \ll F_p$. We anticipate that CTP structures combined with electrical control of the QD emission energy \cite{26} could allow for an ultrafast electrical switching of the QD emission rate in order to either efficiently emit a single photon or store the quantum information in the exciton state.

\begin{acknowledgments}
This work was partially supported by the French ANR P3N DELIGHT. O.Gazzano acknowledges support by the French D\'el\'egation G\'en\'erale de l'Armement. 
\end{acknowledgments}


\end{document}